# Algorithms and Decision-Making in the Public Sector

Karen Levy, Kyla E. Chasalow, and Sarah Riley

Department of Information Science, Cornell University

**Keywords**

algorithms, technology, government, policy, prediction, accountability

**Abstract**

This article surveys the use of algorithmic systems to support decision-making in the public sector. Governments adopt, procure, and use algorithmic systems to support their functions within several contexts—including criminal justice, education, and benefits provision—with important consequences for accountability, privacy, social inequity, and public participation in decision-making. We explore the social implications of municipal algorithmic systems across a variety of stages, including problem formulation, technology acquisition, deployment, and evaluation. We highlight several open questions that require further empirical research.

## 1. INTRODUCTION

High-profile news about algorithms often stems from the private sector. Some stories are complimentary and optimistic, describing algorithms' potential to transform society for the good; others are less generous, focusing on job displacement, threats to privacy and autonomy, or the role of algorithms in promoting misinformation and polarization on social media. These phenomena capture public attention and influence how people perceive algorithms, shaping policy, legislation, and research agendas. Public-sector algorithmic systems reside somewhat outside this spotlight, especially at the local level. These systems influence government functions ranging from the mundane (waste collection) to high-stakes (pretrial judicial decision-making). Yet despite their reach and impact, empirical work on the impact and efficacy of public-sector algorithmic systems remains nascent.

By synthesizing existing empirical and analytic work on these controversial technologies, we aim to provide researchers with a roadmap to guide empirical inquiry into public use of algorithmic systems—particularly at the local level and with a focus on the United States. We highlight several strategic research sites for investigating public-sector algorithms, organizing them by stages in the life cycle of algorithm development and deployment. In practice, the development and adoption of algorithmic systems are iterative processes, unlikely to take place in a set of distinct, ordered steps. The life-cycle approach we propose is an analytic tool rather than a faithful representation of reality. That said, the stages we describe highlight common themes and open questions, while pinpointing decisions and assumptions that are likely to provide empirical detail and analytic leverage for the social scientist.

We proceed as follows. In Section 2, we introduce our inquiry, noting definitional and scope ambiguities that complicate discussions of algorithms and providing background on algorithms in the public sector. In Sections 3 through 8, we highlight key tensions, constraints, and dynamics that emerge across different stages of the life of the public-sector algorithm. Section 3 concerns problem formulation—the identification of a social problem to be addressed and its translation into algorithmic terms. Section 4 considers the role of data—their acquisition, integration, and role as both a resource and a constraint. In Section 5, we turn to how systems are acquired through procurement from commercial vendors or other means. Section 6 considers approaches to governance of public-sector algorithms, from transparency requirements to outright bans. In Section 7, we turn to intersections with frontline work and responses by communities and civic organizations. Section 8 considers how algorithmic systems are maintained, evaluated, reconsidered, improved, or dismantled over time. Section 9 concludes.

## 2. THE PUBLIC-SECTOR ALGORITHM

### 2.1. Algorithms and Algorithmic Systems

"Algorithm" is a charged word today, used to invoke a powerful, sometimes mysterious technology that drives decisions and shapes our on- and offline experiences (Kitchin 2017, Ziewitz 2016). On one level, an algorithm is simply a "set of rules that precisely define a sequence of operations" (Stone 1971, p. 4). For example, machine learning, the subject of much contemporary algorithm talk, relies on training algorithms to learn a pattern between inputs and outputs of interest. In fact, two algorithms are involved here: the training algorithm, which specifies a procedure for learning from data, and a final learned model, which takes in input (e.g., medical details, traffic patterns, arrest data), applies a rule to it, and produces outputs (often, a prediction) (Kleinberg et al. 2018, Lum & Chowdhury 2021). "Algorithm" is sometimes used broadly to refer to either or both.

Other terms further complicate matters, including artificial intelligence; machine learning; automation; and variants of predictive analytics, data science, and statistics. Though it is out of scope here to delve into the finer distinctions among these terms, it is worth noting that they are used somewhat interchangeably in the context of data-intensive technologies—though sometimes with very different connotations. Artificial intelligence, or AI, is an umbrella term for various computing technologies that aim to perform tasks understood to require some form of complex cognition. At times, the phrase may conjure futuristic images of systems with humanlike abilities, perhaps at the cost of attention to current technologies that are incapable of such autonomy but powerful nonetheless (Krafft et al. 2019). Similarly, machine learning, impressive as it may sound, describes what are fundamentally statistical techniques for fitting (sometimes very complicated) models to data. However, because not all algorithms can be classified as machine learning, and even relatively simple algorithms—including explicit, rule-based formulae (Citron 2008)—can play significant roles in policy contexts, some speak instead of automated decision-making systems. This term focuses greater attention on the political function these systems play—to assist or replace human judgments—rather than the technical procedures that underlie it (Molnar & Gill 2018, Spielkamp 2019).

Definitional decisions are more than semantic squabbles. They affect communication among researchers and policy makers, shape public understanding, and determine which systems are subject to regulation (Krafft et al. 2019, Young et al. 2019b). That is, terminology can itself be a substantive and divisive policy issue, not merely a matter of procedural scoping. For example, in 2018, New York City convened an Automated Decision System Task Force—a panel of experts from academia, industry, civil society, and city agencies that was charged with providing recommendations to city government regarding the use and governance of these systems. The panel spent its first several months failing to reach consensus on how broadly to define its object of study (Richardson 2019), questioning whether (for example) the use of an Excel spreadsheet qualified as an "automated decision system." This definitional effort was further complicated by the city's reluctance to provide data on algorithmic systems currently in use, making debates around terminology and scope necessarily abstract (Cahn 2019). The panel's challenges highlight both the difficulty of building consensus on the definition and scope of algorithmic systems and potential barriers to grounding such discussions in current practice.

Further, the terminology of the algorithm carries with it the capacity to create rhetorical distance between policy-makers and policy outcomes—a distance that can be employed to deflect responsibility for unpopular decisions. As Lum & Chowdhury (2021) describe, personnel at Stanford Medical Center were quick to blame a "very complex algorithm" that "clearly didn't work right" when their system for allocating Covid-19 vaccines prioritized high-level administrators over frontline workers. The algorithm had, in fact, worked right—in the sense that it dutifully implemented the rules that decision-makers had determined. When policy outcomes meet public opprobrium, deflection of blame to the algorithm serves a political function, allowing decision-makers to save (at least some) face.

Here, we use algorithm or algorithmic as general shorthand for technologies that rely on machine learning techniques or explicitly programmed rules to inform or execute government actions. As many scholars have emphasized, these technical processes operate within social and institutional contexts, comprising sociotechnical systems (Kitchin 2017, Selbst et al. 2019, Spielkamp 2019). In this vein, social scientific research on public-sector algorithms generally focuses on the political, legal, economic, and social dimensions of algorithmic systems rather than on their computational components alone. We follow suit here.

## 2.2. Roots and Growth of Public Sector Algorithmic Systems

We focus our analysis on local governments in the United States because of the scope of their responsibilities and their high degree of influence on public life. Local governments have been called "federalism's workhorses," as they "lead in the expenditure of funds for domestic purposes and employ the greatest number of people to carry out these goals" (Advis. Comm. Int. Relat. 1982, p. 11). They are the leading provider of services to the public, overseeing public safety, sanitation, parking, and education services that many rely on daily. Given the volume and breadth of services localities administer, they are fertile ground for emerging automated systems. If states provide 50 laboratories of federalism, municipalities provide more than 90,000 (Hogue 2013).

Though the daily workings of local government affect everyone, more vulnerable constituents are most often exposed to and directly impacted by municipal algorithmic systems, often without notice or the choice of opting out. Individuals with the least access to material resources, especially people of color, are more likely to interact with government agencies—both because their marginalized status gives them fewer private options and because the state proactively and disproportionately controls these populations (Barabas 2020, Bridges 2017, Eubanks 2018, Madden et al. 2017). Municipal algorithmic systems thus have disproportionate influence on the lives of the most vulnerable.

Though we focus on the recent growth of public-sector algorithmic systems, these systems did not, of course, appear out of thin air. They are rooted in a long history of government data collection; statistical analysis; and the expanding role of these tools in government administration, policy-making, service delivery, and population management. Their tributaries include foundational data infrastructures—civic registration systems, censuses, surveys, bureaucratic administration—as well as statistical and social scientific methods often developed with close links to government planning and service delivery (Bouk 2015, Breckenridge & Szreter 2012, Didier 2020, Salsburg 2001, Scott 1998).

Broadly, public sector algorithmic systems emerge from a zeitgeist of formalization, rationalization, and automation in government that has been present since at least World War II. The ideal of systematized, quantified government decision-making has roots in operations research, with its ambition to develop neutral, scientific, and objective techniques for analyzing situations and determining optimal courses of action (Harcourt 2018). A key early actor was the RAND Corporation, a think tank founded in 1948 to provide research to the US Air Force. RAND subsequently deployed its methods in cities across the country to "reshape the municipal bureaucracy along rational, scientific lines" rather than the "reactionary muddling through" said to characterize city governments' typical responses to problems (Flood 2010, pp. 111, 116). The mid-twentieth century also saw early forms of automation in the guise of expert systems, meant to automate decision-making by encoding domain expertise into systems of logical rules (Hadden & Feinstein 1989). Related ideas are reflected in the push for evidence-based policy and decision-making (Engstrom et al. 2020, Stevenson 2018) and in the growth of actuarial practices that rely on statistical prediction rather than on clinical judgment developed through experience and training (Dawes et al. 1989, Simon 1988).

If the push for formalized, data-driven decision-making in government is not entirely new, what, then, is distinctive about today's algorithms? A simple answer is their pervasiveness. New algorithmic tools are more "deeply embedded" in government processes than past efforts, "moving the new algorithmic governance tools to the center of the coercive and (re)distributive power of the state" (Engstrom et al. 2020, p. 11). Enabling those tools are increased collection and integration of data, which support an expanding capacity to harness "collective intelligence" on public service delivery and citizens' lives (Vogl et al. 2020). Those data are also processed in new ways: Machine learning techniques are characterized by a focus on developing models for prediction rather than explanation (Breiman 2001, Hofman et al. 2017, Kleinberg et al. 2015), and by doing so without coding rules explicitly, making them harder to interpret and explain (Caruana et al. 2015, Lipton 2018, Selbst & Barocas 2018).

Two other noteworthy features often attributed to algorithmic systems today are scalability and personalization. Scalability refers to the way algorithms can be used to influence "decisions uniformly and comprehensively," replicating decision-making sequences (and any assumptions or flaws they incorporate) across contexts and over time (Brauneis & Goodman 2018, p. 129). Personalization implies an expansion from long-standing uses of data for broad policy planning based on aggregate data and toward uses of data that "categorize citizens, allocate services, and predict behavior" on an individual basis (Dencik et al. 2018, p. 3). The hope is that this will allow better resource allocation, lower costs, and more preventative intervention.

Though the reasons that local governments adopt these systems are situated in the broader histories and current trends, local governments occupy a distinct position, with particular constraints and needs that make algorithmic systems particularly appealing. Though local governments vary considerably in terms of funding, organizational structure, and policy priorities, some of their common features help explain algorithmic system uptake at the local level.

First, local governments have less financial power relative to other levels of government. They rely on a complex and ever-shifting patchwork of funding from state and federal governments—complex in part because amounts change over time, but also because different categories of funding (e.g., categorical grants, block grants) may come with complicated conditions. The proportion of local revenue that comes from intergovernmental transfers has declined steadily for decades (Berman 2019, p. 111; Wildasin 2010). In addition, many states restrict how and how much revenue local governments can raise on their own through measures such as tax and expenditure limitations. These measures, like California's Proposition 13, reduce local budgets and force localities to rely on alternative methods for generating revenue, often through politically unpopular fines and fees. Thus, for all their many responsibilities, local governments face the dual hardship of aid reductions and cumbersome grant conditions. This makes modern systems that promise to accomplish more with fewer resources highly desirable.

In addition to financial stressors, local governments face the pressure of delivering high-stakes and highly visible public services. Localities frequently make critical decisions, for which constituents can trace responsibility to a single office or even individual: Judges decide who to release or detain prior to trial, fire departments prioritize buildings to inspect for code violations, and housing authorities distribute limited rental assistance. Agencies must make difficult decisions in the best interests of their constituents while navigating budgetary and political constraints. Incorporating algorithmic decision-making support may therefore have particular appeal as a way for localities to try to navigate these tasks objectively and accurately.

## 3. FORMULATING THE PROBLEM

### 3.1. Problem Definition

Algorithmic solutions are, ostensibly, brought to bear on problems. But how do governments come to identify problems as in need of algorithmic solutions? The act of recognizing, defining, and specifying the problem that an algorithm is intended to address is itself a deeply political question.

Policy scholars have long characterized certain problems as "wicked"—complex and messy, lacking in clear boundaries or obvious solutions, and subject to conflicting interests. Poverty and climate change are classic examples. Policy-makers often "carve off" tractable components of these systemic problems that are amenable to being "solved" (Churchman 1967). This involves simplifying the wicked problem into one that is easily measured and politically palatable. This is where algorithms come to the fore. As Kaplan (1964, p. 303) put it, "We tend to formulate our problems in such a way as to make it seem that the solutions to those problems demand precisely what we already happen to have at hand."

The process of converting a systemic problem into an algorithmic problem has important implications for what problems are addressed and how they are understood. Some problems do not lend themselves readily to algorithmic solutions—or by making them fit such a solution, the shape of the problem changes considerably. The part of a problem that can be addressed algorithmically may steal political oxygen from nonalgorithmic reforms or palliate concerns that "something must be done" but without addressing root causes (Abebe et al. 2020a)—what Churchman (1967) termed "the taming of the growl" of a wicked problem. For example, Eubanks (2018, p. 197) notes that before cities and states brought algorithmic systems to bear in managing poverty programs, "the nation was asking difficult questions: What is our obligation to each other in conditions of inequality? How do we reward caregiving?" She points to algorithms as "refram[ing] these big political dilemmas as mundane issues of efficiency and systems engineering: How do we best match need to resource? How do we eliminate fraud and divert the ineligible? How do we do the most with the least money?" (Eubanks 2018, pp. 197–98). Not only did algorithmic systems fail to address the root causes of poverty, they narrowed what was seen as being on the table politically and altered the values used to evaluate those options.

Unsurprisingly, problem definition depends in large part on who is doing the defining. Community stakeholders may conceptualize a problem through the lens of an underlying social issue (poverty, homelessness, insufficient public transit). From within government, problems may be formulated as operational inefficiencies, like fraud reduction or cost savings. Alternate framings may or may not be articulated explicitly. The impetus for algorithmic solutions at the local level is also sometimes shaped by other levels of government, a dynamic that Solow-Niederman et al. (2019) term "algorithmic federalism." For example, after the September 11 attacks, the federal government greatly expanded financial support for local law enforcement in an effort to combat terrorism, spurring the development of surveillance programs by local police departments (Crump 2016). Problems may form in part as a function of the resources available to address them.

## 3.2. Problem Formalization

Addressing social problems through algorithmic means requires an additional translational step, whereby constructs are operationalized in formal, measurable terms. Turning a social problem into an algorithmic problem requires explicit formalization of inputs and desired objectives. In so doing, it necessarily creates "a particular lens for understanding what the problem is" (Abebe et al. 2020a, p. 254).

The translational work of problem formulation "do[es] necessary violence to the world that [it] attempt[s] to model, but also provide[s] actionable and novel ways to address complex problems" (Passi & Barocas 2019, p. 9). The need to formalize a problem into algorithmic terms entails trade-offs: Though it can reduce and oversimplify complex problems, it may also aid in precise articulation of general policy aims and make plain different understandings of a construct. Given sufficient transparency and deliberation, stakeholders may be able to leverage the formalization required for algorithmic systems as a means of asking (for instance) what specifically we mean by risk, or by clarifying exactly what factors should determine the order of some allocation. Formalization can, at its best, make vague policy judgments concrete and provide touchpoints for focused legal and political advocacy (Abebe et al. 2020a). However, the process of formalizing a problem in terms amenable to measurement, statistical modeling, and available data involves decisions that, if not considered carefully, can change the nature of the problem and the outcomes that follow.

Consider the development of the Silicon Valley Triage Tool in Santa Clara County, California. The tool uses data from seven county agencies to predict which homeless people are likely to become "high-cost public service user[s] in the next year (Early 2016, Toros & Flaming 2018, p. 119). The goal is to prioritize these people for permanent housing and thereby reduce costs. Crucially, the choice to use risk of incurring high costs as the outcome for this tool is an ethical decision. It may well be a justifiable one: High cost may correlate with high need, and reduced costs may free funding for other people and programs. But the tool could underprioritize young people who cost less now but may cost and suffer more over the long term—and prioritizing high cost may yield inequitable results if women, immigrants, and minorities tend to be lower cost, including for reasons unrelated to their level of need (Harkinson 2016, Obermeyer et al. 2019). Though described as "a fair, objective tool for triage" by its creators, it can be so at best only given a set of decisions about what outcomes to (objectively) predict (Toros & Flaming 2018, p. 118). Conversely, if those decisions are not well-established prior to building such a model, then the model's creators implicitly implement a policy choice through their choice of target variable. (For a different model of allocating housing assistance—using probability of reentry into the shelter system as an outcome—see Kube et al. 2019.)

Similar concerns apply to the choice of variables to use as inputs for a prediction or classification task. Although predictive accuracy is a primary concern here, there has been much recent discussion of how to navigate the potential discriminatory impact of using certain features (even when they contribute to accuracy) and how to then respond to and mitigate these issues (Barocas & Selbst 2016). A related concern is whether choices of input variables have been made under a "deficit-based approach" that relies mainly on data on negative factors—indications of high risk—without also incorporating positive indicators of lower risk (Brown et al. 2019).

The deficit approach and concerns about disparate impact might each be described as a kind of bias. It is often warned that data and models may be biased—but the big question is, in what way? All data are unrepresentative in some way (e.g., relative to some targets or for some variables); all data have limitations (Chasalow & Levy 2021). The task becomes to make those limitations explicit and investigate whether they matter for the goal at hand. For example, it is important to distinguish the matter of selection bias from that of historical bias (Suresh & Guttag 2020). Whereas the former may be resolved by a better data collection scheme, the latter arises even with perfect measurement, when data capture past realities that are in some way undesirable. Those data are biased in the sense that predictions formed from them will tend to replicate those past realities. The very choice to use historical data (and in some sense, all data are historical) may be viewed as a policy choice in that it expresses a wish or at least a willingness to replicate aspects of the past to achieve policy goals. For future possibilities and untried systems, we have no data.

Policy choices further extend to model development. Mulligan & Bamberger (2019) give as an example the use of seismological models for predictive policing (the intuition being that crime propagates like earthquake aftershocks), but not all modeling relies on such intuitive analogies. Regression-based approaches and more complicated machine learning models can perform well while being difficult to interpret (Lipton 2018). In any case, learning a model from data requires choices about how to measure model fit or predictive performance. It is well known that such choices involve trade-offs—for example, between false positives and false negatives (Kleinberg et al. 2017). A growing literature on algorithmic fairness explores ways to impose constraints that require models to achieve different group- or individual-level fairness criteria (Chouldechova & Roth 2018, Mitchell et al. 2021, Mulligan et al. 2019). Ultimately, designers and governments must also make a choice so basic that it may be overlooked: whether or not to use a predictive model at all. At times, even the best model among those developed may have an unacceptably high absolute level of error (Salganik et al. 2020). Moreover, in some cases, slight differences in problem formalization can lead to very different policy outcomes. An example arises in Abebe et al. (2020b) exploring possible allocations of subsidies meant to prevent sudden financial disruptions (income shocks) from causing financial ruin for households. Though the general policy aim (preventing adverse financial outcomes for families) and policy mechanism (providing subsidies) are consistent, slightly different plausible formalizations of the problem can result in entirely reversed orderings of who should receive assistance.

Allocation problems in particular raise another distinction: Although many algorithmic systems target a prediction or classification task where it makes sense to consider measures of error, not all algorithms are formed this way. For example, Lee et al. (2019) build an algorithm to allocate donations for a nonprofit food distribution organization by aggregating the preferences of a variety of stakeholders. In this case, the algorithm is built to reflect stakeholders' choices about what factors to prioritize, and there is not a true correct allocation to compare it to. That is, an algorithm may try to accurately predict a measured outcome (e.g., costs incurred by homeless individuals), but it may also implement a set of policy preferences directly (Lum & Chowdhury 2021).

## 4. ACQUIRING DATA

### 4.1. Data as Resource and Constraint

The life cycle of the government algorithm begins with a problem or task. It also begins with data. Local governments are tremendous generators of data about the daily lives of individuals and communities. These data take many forms (structured versus unstructured, human versus machine generated, among others), may reside in different technological systems, and may or may not be produced with later analysis in mind. Data production is shaped from within and without—by technological infrastructures, by organizational procedures, by laws and politics, and by employee judgment and external contingencies (Ku & Gil-Garcia 2018). Crucially, the data governments collect or can access shape and reflect the kinds of problems governments are willing and able to address. Data enable but also constrain, and this dynamic manifests in multiple ways.

Problem definition often raises a tension between the measurability of a variable and its relationship to the true outcome of interest. Measurability and construct validity are common concerns across empirical social science, where the variables we can measure or access are proxies for true outcomes of interest (Jacobs & Wallach 2021, Singleton & Straits 2005). These considerations are acute in the policy context, where implemented programs and the predictions they make may determine life outcomes for real people. Some of the measures we most care about in terms of life outcomes are difficult to measure and hard to isolate causally, making policy evaluation difficult. Even when extensive high-quality data about people's lives are available, outcomes can be difficult to predict at the individual level (Garip 2020, Salganik et al. 2020). This ought to give us some pause about policies that center individual-level prediction.

Given the complexity of the social world, model builders may be inclined to instead measure internal operational outcomes less subject to chance and noise but more distant from the consequence they really care about. An example arises in Flood's (2010) study of fire department closures in New York City: Because more direct measures of success or failure (fatalities, injuries, property damage) are relatively rare and impacted by factors outside the department's control, modelers decided to use response time—the time between an alarm being pulled and the fire truck's arrival, which could be measured by giving stopwatches to lieutenants—as a proxy (Flood 2010). This was in many ways an imperfect measure: Fire truck arrival does not perfectly map onto success in fighting fires, and uncooperative firefighters were happy to fudge the data. Using an internal outcome measure simplified the model and made it analytically tractable, but at the cost of reliance on questionable assumptions about outcome validity.

Another example arises in Eubanks's (2018) study of Allegheny County's Family Screening Tool, a model used to forecast the risk of child abuse and neglect in order to prioritize interventions by caseworkers. Here, the true outcome of interest is child maltreatment, which is difficult to measure directly. Instead, Allegheny County used two proxy variables as outcomes: community rereferral (whether the county received another report about the child) and placement in foster care (which resulted from actions by the agency itself and the courts). As a result, the Family Screening Tool predicted future behaviors of the community—whose reports of suspected abuse disproportionately targeted Black and biracial families—and of the agency itself, rather than the direct outcome of interest. These types of simplifications are common and often necessary to be able to build a model at all, yet they can introduce bias, room for manipulation, and circularity into predictive systems.

Even if fairly direct measures of outcomes of interest are available, they might not be available about everything or everyone. Government-collected data are often subject to selection bias: Governments collect more data about people who have higher degrees of contact with government programs and institutions (Chouldechova et al. 2018, Dettlaff et al. 2011). Allegheny County's Family Screening Tool relied on data gleaned from public programs: welfare offices, Medicaid, mental health services, and the county jail, among others. However, the county lacked access to data about residents who did not have contact with these programs and institutions—residents who could, for example, afford private health services. The result, in Eubanks's words, is that "the model confuses parenting while poor with poor parenting" and is relatively blind to some segments of the population (Eubanks 2018, p. 158). A predictive model constructed with historical data may, in some cases, improve decision making—but it also risks reproducing or exacerbating the structural inequalities embedded in the system.

## 4.2. Data Integration and Interagency Dynamics

Algorithmic systems often involve coordination across multiple public agencies and are often accompanied by a push for data integration—that is, for the linking of data from departments and services within or across municipalities, or of data from the public and private sector, to create a more integrated view of government operations (Malomo & Sena 2017). Data integration may be viewed as a means to more accurate prediction; earlier, more targeted preventative measures; or better insight into the workings of government services (Dencik et al. 2018, Toros & Flaming 2018, Young et al. 2019a). Related efforts can be found in proposals for smart cities, enabled through the deployment of sensors (Green 2019) and open data portals that allow members of the public and the private sector to draw on government data (Schrock 2016). The vision is one of more efficient, effective, and connected government.

Data integration is not easy. Public sector data are often fragmented, constrained by organizational and technological infrastructures that limit the ability to share data across organizations or services. Calculating a pretrial risk score, for instance, requires personal and demographic information (e.g., age, employment and housing status), charge information, and criminal history, which may come from different agencies and databases. Risk assessments therefore require ongoing coordination and cooperation to deploy and may be hindered by agencies having different constraints and infrastructures not built with such uses in mind. Malomo & Sena (2017) describe how the "structural data silos" that result from the fragmentation of public-sector activity and a lack of common data standards mean that data sharing can take considerable legal and technological effort. In the case of public–private partnerships, resistance to integration may stem from companies' fear of losing competitive advantage if they share proprietary data (Young et al. 2019a).

But these barriers may also serve as protections (Hartzog & Selinger 2015, Surden 2007). Data integration raises the prospect of far-reaching surveillance and privacy violations, of coordinated decision-making systems that connect previously disparate areas of people's lives, and of blurring between the public and private sector (Nissenbaum 2010, Ohm 2009). For example, in Pasco County, Florida, the Pasco County School Board and the state Department of Child and Families each collect data on students to support educational goals and the health and well-being of children and families. These data have also, however, been combined by the Pasco Sheriff's Office, together with its own records, to "identify juveniles who are at-risk of becoming prolific offenders" and maintain a roster of students it believes are "most at-risk to fall into a life of crime" (Bedi & McGrory 2020). In this case, data integration has enabled controversial uses quite different from those for which the data were collected originally.

Data integration is all the more concerning given inequalities in data availability. Earlier, in the context of Eubanks's child welfare case, we referred to harms that stem from being overrepresented in administrative data, but the benefits and harms of integrated public-sector data do not track cleanly with exclusion and inclusion in data sets (Lerman 2013). Inclusion can yield surveillance and service—often at the same time. Exclusion can reflect and reinforce power and resources (e.g., families able to afford private health services receive less scrutiny from welfare agencies) or disadvantage and marginalization (e.g., people with fewer means of reporting problems or advocating for their needs) (Crawford 2013). Researchers face these dynamics, too. As Johnson & Rostain (2020) argue, public sector data may serve as both "surveillance" and "spotlight," the latter a metaphor for the potential to use it to monitor institutions and expose inequities and rights violations.

## 5. PROCURING THE SYSTEM

Problem definition, data acquisition, building the system—often these do not happen entirely in-house. Instead, many algorithmic tools are purchased from private vendors that build and license them to municipalities. Procurement is a routine part of how governments acquire technologies and is often necessary given their limited access to in-house expertise and resources. Typically, agencies issue a request for proposal to solicit bids from vendors. In theory, requests for proposal enable agencies to select the best solution to a particular problem and increase government transparency, though this does not always happen in practice. In fact, procurement processes pose challenges for governments' ability to retain control over how they provide services and govern. Algorithm procurement raises questions about responsibility (Whose is it?), discretion in design and decision-making (Who has it?), and policy-making (Who does it? Who knows about it?). At stake, Brauneis & Goodman (2018, p. 109) write, is the potential "corporate capture of public power," leaving "the government, which alone is accountable to the public…hollowed out, dumb and dark."

### 5.1. The Procurement Mindset

In the United States, local, state, and federal government agencies have a degree of freedom in how they implement their legally mandated functions. When governments' actions rise to the level of substantive policy decisions, administrative laws often require them to follow procedures intended to promote values such as transparency, accountability, public involvement, and reasoned decision-making (Adams 2018, Davidson 2016, Mulligan & Bamberger 2019). Such procedures include requiring agencies to keep records of the deliberations and evidence used to justify policy decisions, records that may later be subject to judicial review. Government procurement, though subject to regulation intended to ensure fair competition among vendors, is often exempt from these administrative requirements. Mulligan & Bamberger (2019) warn that current algorithm acquisition practices create the potential for substantive policy choices to be embedded in contracting decisions and left to third-party developers. The "procurement mindset," they argue, positions algorithm acquisition as a matter of internal management. It views algorithmic systems as providing a new mechanism to fulfill an existing, "well-defined function" rather than as methods that might change how "an institution makes sense of and executes its mission" (Mulligan & Bamberger 2019, p. 787).

Recall the example of the New York City task force, caught on the question of whether even Excel fell within the umbrella of "automated decision systems." To be sure, some government systems that rely on algorithms are truly part of internal management—a cybersecurity system, for example—and the procurement mindset may be appropriate there. This does not give such systems a full pass from scrutiny, as even seemingly mundane decisions about technical infrastructure can be impactful (Kelion 2020). But the procurement mindset may be especially problematic for public-facing systems that affect people's rights and obligations, constrain agency discretion, and involve choices about how to translate general legal mandates to specific implementations and enforcement rules. A key task then becomes determining which systems are inward-facing and which are public-facing enough that they should require more transparency and public justification (Mulligan & Bamberger 2019). We do not draw that line here, but the first step is to delineate the kinds of policy choices that could make an algorithmic system more than just a matter of implementing preexisting goals and policies.

The procurement mindset critique implies that, in contrast to decisions that merely implement a well-defined procedure, policy decisions affect rights, values, and social outcomes, involve ethical trade-offs, or determine the allocation of valuable resources. Procurement raises the important questions of which stakeholders hold power to define and formalize an algorithmic problem and which processes shape this participation. A wave of recent research has explored what role participatory decision-making processes might play to ensure that public priorities are well reflected in the formulation of the algorithmic problem from the outset (Lee et al. 2019, Sloane et al. 2020, Young et al. 2019c). Yet procurement may pose a barrier to that participation. When the process of building an algorithmic system is outsourced to vendors or professional analysts, policy-makers may come to view the formalization of a model into algorithmic terms as the province of quantitative experts. But once authority is ceded to quantitative experts, those experts may not view their role as evaluating whether the problem definition is appropriate, whether the assumptions it reflects hold in reality, or what nontechnical implementation components (like due process protections) need to accompany the technical tool (Goodman 2019). As Flood (2010, p. 219) writes, though RAND's quantitative experts

> were ultimately responsible for their flawed assumptions [reflected in their firehouse closure model], it's a bit like expecting a cement-truck driver to ponder whether building new roads is the best way to alleviate the nation's transportation problems—thinking about that isn't his job, and thinking it out loud might actually get him fired.

## 5.2. Alternative Paths to Procurement

Procurement processes vary. Governments can be involved to different degrees, with different kinds of contractual ties to companies, and with more or less off-the-shelf or customized systems. Though often developed by private companies, algorithms can also be procured through open competition. In 2017, Boston Public Schools announced the $15,000 Transportation Challenge to design an algorithm to reallocate its bus routes and school start times while meeting a set of stated goals (Goodman 2019, Whittaker et al. 2018). Those goals were for the algorithm to improve route efficiency and redress inequities in the existing distribution of school start times, which disadvantaged minority and poorer families, while also meeting a set of constraints motivated by health and safety considerations. The algorithm built by the winners, a team of Massachusetts Institute of Technology researchers, met these goals (Bertsimas et al. 2019). Yet this was not enough to prevent fierce community backlash, particularly against the disruptiveness of the algorithm to existing routes and times, leading the district to partially abandon the project.

Goodman (2019) argues that the problems that led to the algorithm's partial demise stem from insufficient engagement between the public and the model, and that this issue was already apparent in how the competition was framed, without any emphasis on sociopolitical implementation or communicating with the public about how the algorithm worked. Though the school district recognized the equity issues at stake and engaged the public about "what they wanted out of bus transportation and school starts in the abstract," she writes, "there was almost no engagement with the model itself, insufficient transparency about the algorithm's tradeoffs, and no opportunity to adjust it" (Goodman 2019, p. 3). More direct communication about trade-offs between increasing equity and minimizing disruption might have led to more public acceptance or to an adjusted algorithm that also prioritized stability. Although procurement by contest inherently delegates some decisions to the competitors, the powers of task definition, evaluation, adoption, and public presentation remain in government hands.

Despite its failures, the Boston busing algorithm offers an example in which a government obtained an algorithm to help with a complex task while openly stating its policy aims, making some attempts at public engagement, and retaining close control over the evaluation and implementation of the algorithm. It is exactly these things that are often lacking from procurement processes, especially when they occur through private vendors—part of a larger set of concerns about the evolving balance of power between governments and technology companies (Pasquale 2015, Zuboff 2019). Recently, describing the case of the Oakland Police Department's use of ShotSpotter (a widely deployed technology for detecting the sound of gunfire), Mulligan (quoted in Kaye 2021) painted a stark picture:

> They [the city government] actually don't get the data about movement patterns in their own city…and so in some ways they end up becoming hostage to companies that now know more about the urban infrastructure than the city may because they have the raw data … and the city is just left with a set of reports.

It is this kind of dynamic that Brauneis & Goodman (2018, p. 109) have in mind when they write that procurement threatens to leave governments "hollowed out, dumb and dark."

## 6. GOVERNING THE TECHNOLOGY

### 6.1. Transparency and Assessment

The image of the black-boxed algorithm, or of procurement without public scrutiny, begets calls for openness as a road to critical evaluation, democratic process, and accountability. However, meaningful transparency has proved difficult to define and realize. Transparency can have many objects, including source code and modeling assumptions, data set details, internal deliberations, reasoning for actions actually taken, contracts, and budgeting. It makes sense to think of transparency as a multidimensional goal, measured in degrees, for which different dimensions raise different technical, legal, and social stakes (Burrell 2016, Coglianese & Lehr 2019, Fink 2018, Levy & Johns 2016, Pozen 2018, Pozen & Schudson 2018).

Amid this multiplicity, one broad but useful distinction is between more proactive and reactive forms of transparency—the former occurring more by design, legal obligation, and voluntary disclosure, the latter in response to ad hoc information requests (ben-Aaron et al. 2017, Bloch-Wehba 2020). Reactive transparency may sometimes be too little too late. Mulligan & Bamberger (2019) emphasize the need for interventions at the level of design and procedure and note that ex post transparency is insufficient for challenging policy decisions embedded in systems already in use. When governments procure algorithmic systems from vendors, they are sometimes left in the dark about the details of how systems work. Presented with open records requests, governments may be unable to fulfill them because they do not possess the information themselves. Vendors may claim trade secret exemptions to sharing data, software, and design processes, and government contracts may allow them to do so (Wexler 2018). Agencies themselves can also have an interest in secrecy and sometimes interpret nondisclosure agreements and trade secret exemptions broadly (Brauneis & Goodman 2018).

Information requests, though reactive, remain an important mechanism of public exposure and a way that scholars study transparency empirically. Ben-Aaron et al. (2017) use information requests to local governments to study whether governments are more likely to fulfill requests when told that other counties have already complied. Brauneis & Goodman (2018) use open record requests to catalog algorithms used in pretrial risk assessment, child welfare, predictive policing, and teacher evaluation and thereby reveal gaps in governments' knowledge of their own systems. Such transparency research provides not only information about algorithms and government procedures but also an indication of how willing or able governments are to provide that information in the first place.

Alternatively, proactive transparency efforts focus on requiring governments to produce or release information on system design, acquisition, performance, and projected impacts. In this vein, some policy-makers have turned to documentation requirements in environmental policy making for inspiration. Many proposals for algorithmic impact assessments (or AIAs) call for a mandatory analysis to be completed before a system is acquired or deployed. These assessments can include periods for public comment and response; disclosure of key details about purpose and implementation; discussion of alternative means of achieving policy goals; and prospective evaluation of potential concerns about accuracy, bias, and impact on stakeholder groups (Reisman et al. 2018, Selbst 2017). Similarly, after a system is in use, systems may be subjected to periodic mandatory auditing or other forms of evaluation and testing, to be conducted either internally or by third-party researchers or oversight boards (Ada Lovelace Inst. 2020). These ongoing evaluations focus most commonly on detection of bias, though they may also inspect other qualities, like reliability or compliance with other policy mandates (Kroll et al. 2016).

Empirical data on the effectiveness of AIAs in practice is relatively scant, and policies mandating them are only just emerging (e.g., McKelvey & MacDonald 2019). At best, AIAs might compel proactive reflection on key concerns, build capacity within local governments, and provide a way into the policy process for affected communities. At worst, they might become little more than a checkbox—pro forma paperwork that does little to promote accountability (Metcalf et al. 2021,

Waldman 2019).

## 6.2. Budgeting

Budget processes may also play a role in promoting democratic decision-making about algorithmic systems. In most localities, budget items on the scale of citywide algorithmic systems require city council approval. In theory, budgeting acts as a check, ensuring that public funds are allocated appropriately and giving council members and constituents the opportunity to ask questions about the technology and its expected impacts. That said, budgets are no guaranteed safeguard. For example, when localities receive funds from other sources (i.e., state or federal agencies), they can sometimes skate through the budget process or evade it altogether. In other cases, public notice or comment periods are so brief and poorly publicized that they serve no practical purpose.

Both of these dynamics have played out in Oakland, California. In 2010, the city council approved requests from the police and fire departments to build a Domain Awareness Center (DAC) using federal funds originally intended for antiterrorism efforts. The requests drew little public attention. The DAC aggregated existing data sources, creating a "hub for analyzing video and sensor feeds" (Wood 2013) for the purpose of keeping the Port of Oakland and surrounding area safe. However, when, in 2013, the police and fire departments requested permission to receive a second federal grant to hugely expand the DAC, the public took notice (Winston 2014). This time, an investigation by the nonprofit Oakland Privacy Working Group uncovered that the procurement of the DAC violated a prohibition on contracting with entities involved in nuclear weapons work. Separately, a routine budget conversation revealed the program's full scope and caught public attention. Together, these events catalyzed a movement at a time when residents had been primed by the recent Edward Snowden leaks to react to surveillance programs with skepticism and concern. In response to community pressure, the city council passed a resolution scaling back the program and created a citizen task force (later made permanent) to advise on surveillance technologies. The case demonstrates how budgeting processes can—but do not always—facilitate public oversight.

## 6.3. State Mandates

In the United States, state-level mandates also govern local use of algorithmic systems. The implementation details of state mandates can have profound consequences on the lives of constituents and are often underexamined. These mandates can cede or exert control over different aspects of the implementation process, choices that introduce a tension between global and local understandings of fairness, accuracy, and bias and between centralized and decentralized policy administration (Solow-Niederman et al. 2019).

California's Money Bail Reform Act (SB10) and Kentucky's Public Safety and Offender and Accountability Act (HB463), both intended to govern criminal justice risk assessment practices, illustrate these dynamics. Kentucky mandated that presentence investigations include risk assessment results and that judges incorporate the results in pretrial decision-making. By instating a statewide assessment, Kentucky targeted a version of fairness akin to standardization: An individual could travel anywhere in the state and expect the same risk instrument output. Yet uniform outcomes could be viewed as unfair if different risk factors (e.g., age, criminal history) have different degrees of predictive power across jurisdictions. In contrast, California attempted to reconcile this tension by allowing localities to choose their own risk assessment while also granting oversight authority to the state's Judicial Council. In an effort to strike a balance, California gave the council and localities the power (and responsibility) of validating each risk assessment but created confusion and additional administrative burden (Solow-Niederman et al. 2019). Comparing these two mandates demonstrates how states pursue a complex, and sometimes incompatible, set of goals while also balancing political and budgetary constraints. Without clarifying the state–local tensions and trade-offs that arise under the mandates, states "risk encoding these understandings implicitly, in ways that are opaque and may resist democratic accountability" (Solow-Niederman et al. 2019, p. 724).

### 6.4. Restrictions, Moratoria, and Bans

Beyond transparency requirements, impact assessments, budget measures, and state mandates, local governments may regulate specific algorithmic systems directly. For example, Spivack & Garvie (2020) taxonomize the measures that states and localities have taken with regard to face recognition technologies. The strongest measures ban acquisition and use of systems by the government altogether—or even by both government and private companies. Several local governments have taken such measures (San Francisco, Berkeley, and Boston, among others). Other cities have approached the problem via moratoria that suspend acquisition and use for a set period of time or until further study is complete. Still other governments take more limited approaches, permitting a technology to be used but imposing conditions on its use. As Spivack & Garvie (2020) detail in the face recognition context, these restrictions run the gamut from prohibiting integration with body-worn cameras to limiting access (e.g., permitting access only for investigation of violent felonies). Balancing between purpose- and use-based restrictions on one hand, and wholesale rejections of the technology on the other, is politically controversial: Advocates differ in their assessments of the political palatability and effectiveness of each approach. There is little empirical evidence as to long-term impacts of different strategies.

## 7. USING AND RESPONDING TO THE SYSTEM

Municipal algorithmic systems operate in complex, dynamic environments. They influence and are influenced by multiple agencies and actors, each with their own goals and interests, making it difficult to predict the impacts of a system prior to implementation. Once deployed, additional actors enter the picture, shaping how the deployed algorithm functions in practice—and sometimes pushing back.

It is possible to some extent to anticipate these human–algorithm interactions and to make careful design choices concerning how algorithms are situated within bureaucracies and work environments (Bansal et al. 2020, Dietvorst et al. 2016, Veale et al. 2018, Yang et al. 2020). One important aspect of system design involves not just what an algorithm predicts but how those predictions are presented to the people who use them. Systems are often designed to translate probabilities into user-friendly scales, scores, and categories, but this comes with hazards. If, for example, judges using predictions of pretrial flight risk to make bail decisions are given only risk scores, without being told what probability of reappearing in court a high or low risk score actually corresponds to, they may make inaccurate assumptions about what scores mean. In one validation study of such an algorithm, approximately 90% of defendants earning the lowest risk score of 1 and 70% of those earning the highest risk score of 6 appeared in court (Brauneis & Goodman 2018). Here, the scores 1 and 6 appear to present a more extreme difference than those percentages do—a common hazard when using scores and rankings (Espeland & Sauder 2007). In this way, interpretation and presentation decisions built into the design of an algorithmic system shape its efficacy and impact.

### 7.1. Frontline Worker Responses

Perhaps a still more fundamental decision in the design of algorithmic systems is whether and how to involve frontline workers in the first place. A common distinction here is between full automation and recommender systems (Coglianese & Lehr 2019, Veale & Brass 2019, Yeung 2017). Algorithmic decision systems are often portrayed as displacing human decision-makers, but many systems are not meant to fully replace workers (Kellogg et al. 2020). Instead, they provide decision support (recommendations, predictions, scores), while workers retain the ability to adjust or override the final decisions suggested by the system. On the other hand, some research has demonstrated automation bias, in which workers tend to defer to system output even when they ostensibly have room for discretion (Citron 2008, Skitka et al. 1999).

Research on government worker discretion—whether it exists, the forms it takes, its advantages, and its dangers—has a long history that predates automated decision systems (Bovens & Zouridis 2002). This discretion arises because "street-level bureaucrats often work in situations too complicated to reduce to programmatic formats…[they] have discretion because the accepted definitions of their tasks call for sensitive observation and judgement, which are not reducible to programmed formats" (Lipsky 1980, p. 15). Technological advances have not eliminated this feature of public work, and it is important to understand how workers and machines interact to generate a decision rather than the statistical output of the algorithmic system alone (Guay & Parent 2018, Veale & Brass, Vogl 2020, Vogl et al. 2020). In the criminal justice context, for instance, some research demonstrates significant heterogeneity in compliance with risk assessment recommendations along dimensions of age (Stevenson & Doleac 2019), race (Albright 2019), and income (Skeem et al. 2020). Frontline workers may also actively mitigate system bias by overriding machine decisions (Albright 2019, Stevenson & Doleac 2019). For this reason, some researchers have cautioned against total automation, demonstrating that humans are superior predictors in some contexts (e.g., De-Arteaga et al. 2020). Although agencies deploy algorithmic systems in part for their precision and objectivity, having a human in the loop can cut both ways.

Beyond override decisions, municipal algorithmic systems change workers' experiences—for better and worse—and the ways workers respond to these changes can influence system performance and lead to unforeseen consequences. Brayne & Christin (2020) explore this phenomenon in police departments and criminal courts. They document "strategies of resistance" that police officers and judges have employed in response to surveillance and deskilling. Judges, for instance, worried that administrators and defense attorneys would use risk assessments to assess productivity and performance by directly comparing their sentencing decisions and incarceration rates, thereby jeopardizing their professional autonomy. They also argued that risk scores failed to provide insights beyond what their professional expertise already provided. In response to threats like these, judges and officers employed two strategies: ignoring recommendations and obfuscating data. Police officers, for example, sometimes disregarded the geographic recommendations generated by predictive technologies or even used the technology against itself—deliberately tampering with car antennae to disrupt recording equipment and prevent management from hearing their conversations. Other research on police use of algorithmic systems also demonstrates that officers use systems in many off-label ways that can impair system function and violate assumptions about accuracy (Ehrenkranz 2019, Garvie 2019). Discretion, then, is a double-edged sword: It remains a crucial part of how algorithmic systems are implemented, but bias, pushback, and even system tampering can shape how systems function.

## 7.2. Public Responses

Emerging public opinion research paints a murky and somewhat fractured picture of Americans' attitudes toward algorithmic systems. Survey data indicate a split in public opinion on whether high-stakes automated decisions are effective, fair, or acceptable (Smith 2018). There is also significant disagreement about where to delegate supervisory responsibility for automated systems. In one national survey, only 26% of respondents had confidence in the ability of state governments to manage AI. Confidence in the federal government was roughly the same. In comparison, 41% had confidence in the ability of technology companies to manage AI (Zhang & Dafoe 2019). Perhaps public trust depends on the application: In another survey, 56% of US adults trusted law enforcement agencies to use face recognition software responsibly, while only 36% trusted technology companies to do so (Smith 2019).

However, general trends in public opinion may serve only a limited role in guiding policy action (Brown et al. 2019). Whatever public opinion toward algorithmic systems is in general, a growing number of people and organizations are engaged in efforts to counteract the potential threat these systems pose to their communities. Their goals and methods are diverse. Legal advocates challenge algorithmic decision-making systems on a variety of due–process–related grounds in both criminal and civil contexts—challenging, for example, the use of risk assessment in sentencing without sufficient transparency to allow the defendant to challenge its validity (Wexler 2018) and the deprivation of housing and public benefits on the basis of automated decisions (Gilman 2020, Richardson et al. 2019).

Some groups operate within local government, performing an advisory and oversight role. The Oakland Privacy Commission, for instance, is a citizen-led body that advises the city on how to uphold the community's privacy rights, and Pittsburgh recently created a task force of its own meant to "ensure algorithmic accountability and equity for all residents" (Institute for Cyber Law, Policy, and Security 2021). Other groups exert pressure on local governments from the outside. Examples include Eye on Surveillance, a grassroots organization in New Orleans that works to stop the city's expansion of surveillance tools by organizing public protests and campaigning for anti-surveillance laws, and MediaJustice, an Oakland nonprofit that tackles digital surveillance and high-tech policing and prisons. These diverse public responses mirror long-standing variety in approaches to policy reform. The impacts and trade-offs of different approaches to algorithm reform and opposition remain underexplored, and little empirical evidence on the relative efficacy of the various approaches exists.

## 8. EVALUATING, MAINTAINING, OR DISMANTLING THE SYSTEM

The moment at which a system is introduced is when it is most likely to command attention. But sustained scrutiny is more difficult. As time goes on, some algorithmic tools are likely to fade into taken-for-granted parts of the bureaucratic infrastructure. But a number of key social dynamics emerge only after a system has been in place for a while; these too provide important arenas of inquiry for the social researcher. Notable among these are how a system is evaluated, maintained, or course-corrected over time—if these processes occur at all—and how systems are put to new uses other than those originally envisioned.

### 8.1. Evaluation, Maintenance, and Shift

Evaluating whether an algorithmic system is achieving its intended effects is challenging and context-specific. While evaluating model performance on held-out test data is a routine part of model development, evaluating whether a model continues to perform well on new inputs once deployed introduces further difficulties. Evaluating whether that model is helping a government to achieve its policy goals broadens the scope further still. It requires contending not just with predictive accuracy but with how the system functions in practice.

Algorithms are sometimes promoted as a route to more adaptive policy, preventing it from growing stale and outdated (Mormann 2021). Though at present many public-sector algorithms are not sophisticated enough to adapt in real time to incoming data, there is at least the potential to retrain models periodically on incoming data. But adaptive or not, algorithms are not self-maintaining and exist in dynamic environments. Evaluation over time becomes critical, lest the algorithms themselves become stale and outdated.

Within machine learning, the issue of systems developed in one context but applied in another is often discussed under the heading of "shift" (Quiñonero-Candela 2009, Rabanser et al. 2019). Shift occurs when the real-world conditions under which a model was developed are different then conditions in which it is deployed. Most simply, shift can occur as the world changes over time. This change can reflect gradual drift, sudden external shocks (the Covid-19 pandemic is a case in point), or even governments' own actions. For example, a municipality may adopt policies that affect outcomes that it is also trying to predict. If it does not then update its model with data on the new environment, its predictions may remain blind to the impacts of these changes, creating "zombie predictions" (Koepke & Robinson 2018). Similar issues can occur when a system is trained on data from one population but is then exported to a different population (e.g., to a different municipality) in which conditions differ (Buolamwini & Gebru 2018, Gama et al. 2014, Hamilton 2014, Koh et al. 2020).

Shift is not merely a technical matter; it is also a legal and bureaucratic one. Without procedures and resources in place to support validation studies that test for bias and compare predictions to actual outcomes, shift may go undetected and unaddressed (Koh et al. 2020). In nonalgorithmic contexts, government agencies are no strangers to the need for repeated evaluation. A common legal approach to such issues is the use of sunset clauses or expiration dates by which a program must be revisited or reauthorized (Mormann 2021). These sorts of mechanisms build in purposeful friction and can provide natural cut-points to ensure periodic reevaluation and public oversight. In the algorithmic context, procedures for creating such friction are less developed, but the stakes are high. Automated enforcement tools "are rarely turnkey systems[.]…If data rooted in historical enforcement patterns are unreflectively used to train models and efforts to update those models are ad hoc, enforcement efforts risk…fighting the last war instead of addressing new [problems]" (Engstrom 2020, pp. 25–26).

## 8.2. Feedback

Another set of complications for evaluating algorithms arises when change emerges from the algorithm itself. Social scientific research on performativity provides a useful analog here: The very existence of a system meant to model and measure a given phenomenon can itself shape that phenomenon, bringing human behavior into line with what the system would predict (Callon 1998, MacKenzie 2008).

Feedback loops occur when algorithm outputs are used to make decisions and the outcomes of those decisions are fed back into subsequent training data, reinforcing existing patterns and biases. Bias in arrest data used to train predictive policing algorithms has such an effect: Arrest data for drug offenses overrepresent Black neighborhoods, leading predictions to target those neighborhoods for patrols, resulting in more arrests and thus more overrepresentation (Ensign et al. 2018, Lum & Isaac 2016). Conversely, the model may not be informed about drug use in other neighborhoods where police are not making arrests and fail to target those areas, contributing to cycles of unbalanced policing. Other feedback dynamics include the ways human behaviors may change in response to a model's existence. As Goodhart's Law poses, outcome measures become less useful when actors begin to target them (Abebe et al. 2020a, Bambauer & Zarsky 2018): Schools teach to the test, colleges game rankings (Espeland & Sauder 2007), and job applicants learn what signals to include on their resumes to advance through automated screening (Weed 2021). Such forms of feedback can become sites of contention as those affected by a system try to adapt to and anticipate it while the system's creators try to ensure that it works as intended.

### 8.3. Failure and Creep

How do algorithmic systems fail? We have already alluded to various phenomena that might be called forms of failure: bans, moratoria, legal challenges, shift, feedback loops, and public backlash. An algorithm may also simply fail to achieve its intended effects (e.g., O'Brien et al. 2018). When systems are procured from private vendors, the potential for vendors to "go bankrupt, shift their focus or switch business models" raises additional avenues for failure, as occurred in Alphabet's aborted development of a "smart" waterfront in Toronto (Valverde & Flynn 2020). All suggest potential sites for studying how things can go wrong, who is responsible when it does, and what happens next.

The counterpart to failure is, in a sense, expansion. The life cycle of an algorithm does not necessarily begin or end with a single, well-articulated task. Function creep is a polysemous concept often used to describe how the uses of public-sector technologies may expand beyond their original purposes, particularly when such expansions happen without open debate or acknowledgment (Koops 2021). The introduction of one technology may spur new projects not originally planned or lower the barrier to implementing additional uses of data, tools, funding, and other bureaucratic resources now in place. Of course, such change and expansion can be a form of innovation. But to speak of creep is usually to voice concern and to call attention to hidden and unaccountable change. Analytically, the caveat that function creep poses for our picture of the algorithmic life cycle is that the initial task, problem, or context for which a data set or algorithm is created and deployed may not be all it is used for.

## 9. CONCLUSION: ALGORITHM AS CHALLENGE AND OPPORTUNITY

Public-sector algorithmic systems raise daunting questions about how to ensure government transparency, accountability, and control over their own systems—but they also provide opportunities for transparency and accountability as well as more careful design and explicit problem formulation. Recall the case of the Boston busing algorithm intended to improve efficiency and equity but opposed for its disruption to existing schedules. That algorithm proved to be a mobilizing target for public backlash, prompting slogans such as "families over algorithms" (Goodman 2019). But it is not clear that the algorithm was really the problem. Goodman describes it as a case of "algorithmic scapegoating," in which the algorithm stood in for substantive issues around equity and disruptive change that were really at stake (though potentially more contentious to discuss) and might well have been at stake even without an algorithm in the picture. The tragedy of the case is that the algorithm could have provided the flexibility to involve the public in choosing among multiple trade-offs. If implemented, it might have created a more equitable system than what existed originally.

The duality between challenge and opportunity is a recurring theme in the algorithmic life cycle. In this article, we have traced it from problem formulation, where algorithms can make goals and trade-offs explicit but may also tame wicked problems without addressing root causes; through data acquisition and procurement, where data integration and public–private partnerships expand government resources but come with dangers; and finally to governance, deployment, and evaluation, where algorithms raise legitimate fears but can also become scapegoats. The algorithmic life cycle we present here represents not so much a single lifeline as a collage formed from many arenas, playing out across "federalism's workhorses." It brings into focus important sites for social research. For researchers, the challenge is to make sense of this space amid ambiguous scope, much variety, and high stakes—particularly at the local level, where many of the algorithms that directly impact public life reside. The opportunity is to better understand not just a technology but society and its algorithm-mediated future.

## DISCLOSURE STATEMENT

The authors are not aware of any affiliations, memberships, funding, or financial holdings that might be perceived as affecting the objectivity of this review.

## ACKNOWLEDGMENTS

We are grateful for support from the John D. and Catherine T. MacArthur Foundation (all authors) and the Microsoft Ada Lovelace Fellowship (Sarah Riley).